# Crystal Structure, Magnetism, and Electronic Properties of New Rare-Earth-Free Ferromagnetic MnPt$_5$As

*Xin Gui, Weiwei Xie\**

Department of Chemistry, Louisiana State University, Baton Rouge, LA, 70803, USA

Address correspondence to E-mail: weiweix@lsu.edu; weiwei.xie@rutgers.edu

## ABSTRACT

The design and synthesis of targeted functional materials have been a long-term goal for material scientists. Although a universal design strategy is difficult to generate for all types of materials, however, it is still helpful for a typical family of materials to have such design rules. Herein, we incorporated several significant chemical and physical factors regarding magnetism, such as structure type, atom distance, spin-orbit coupling, and successfully synthesized a new rare-earth-free ferromagnet, MnPt$_5$As, for the first time. MnPt$_5$As can be prepared by using high-temperature pellet methods. According to X-ray diffraction results, MnPt$_5$As crystallizes in a tetragonal unit cell with the space group $P4/mmm$ (Pearson symbol $tP7$). Magnetic measurements on MnPt$_5$As confirm ferromagnetism in this phase with a Curie temperature of ~301 K and a saturated moment of 3.5 $\mu_B$ per formula. Evaluation by applying the Stoner Criterion also indicates that MnPt$_5$As is susceptible to ferromagnetism. Electronic structure calculations using the WIEN2k program with local spin density approximation imply that the spontaneous magnetization of this phase arises primarily from the hybridization of $d$ orbitals on both Mn and Pt atoms. The theoretical assessments are consistent with the experimental results. Moreover, the spin-orbit coupling effects heavily influence on magnetic moments in MnPt$_5$As. MnPt$_5$As is the first high-performance magnetic material in this structure type. The discovery of MnPt$_5$As offers a platform to study the interplay between magnetism and structure.

## *Introduction*

How one can easily approach new ferromagnetic materials (e.g. a practical design strategy), especially those with high transition temperatures and saturated moments, is always a demanding necessity for materials synthetic scientists. Meanwhile, the exploration of ferromagnetic intermetallic materials without strategically vulnerable rare-earth elements is critical for information technology applications, such as magnetic and magnetoelastic sensors,[1] hard-disk drives,[2] spintronics,[3–5] and biomedical devices.[6] Most research focus on using magnetically active $3d$ transition metals to replace rare-earth elements, such as Fe, and Co.[7–10] Some of the 3d elements are ferromagnetic ordered above room temperature (e.g. Fe, Co, Ni).[11] When alloyed with other elements, the magnetic and mechanic properties can be tuned accordingly.[12,13] AlNiCo-type magnetic systems attracted attention for their temperature stability and mechanical properties. AlNiCo has a relatively large coercivity primarily due to shape anisotropy of nano-sized particles rather than the magneto-crystalline anisotropy, which is typical for rare-earth based magnets.[14–18] Later, quite a few Mn-based alloys were discovered to exhibit ferromagnetism, including Mn-B, Mn-Ga, Mn-Ge, Mn-Sb and Mn-As.[19] Particularly, Mn-Al[20–26] and hexagonal NiAs-type Mn-Bi alloys[27–32] have potential due to their large magnetic crystalline anisotropy.

As reported, Mn-Mn distances significantly affected how magnetic moments ordered in some systems, such as $RMn_2Ge_2$ and $RMn_6Ge_6$ (R = heavy-lanthanide elements or Y).[33-35] Moreover, the distribution pattern of Mn atoms in AMnX-based compounds (A = Li, Ni, Cu or La; X = As or Sb) are showing similar behaviors and also related to the saturated moment per Mn atom.[36] In such systems as mentioned, usually ferromagnetic ordering and high saturated magnetic moment are preferred by longer Mn-Mn distance while shorter Mn-Mn distance favors antiferromagnetic interactions. With this intriguing feature of Mn atoms, our candidate system is tetragonal $MgPt_5As$ structure, which can be considered anti-format of one of the most well-known heavy fermion superconductors, $CeCoIn_5$.[37] Among all four atomic sites in $MgPt_5As$-type structure, it is easily to be aware that if Mn atom can be placed onto Mg site in which Mg/As are distributed around 4.6 Å, the neighboring Mn-Mn distance can reach over 4 Å, due to the periodicity of unit cells and the change of atomic radii, which is comparable as the ferromagnetic cases in ref. 36. Furthermore, with the electronic configuration of $[Ar]3d^54s^2$, Mn is well-known to provide high magnetic moments in intermetallics. The spin-orbit coupling (SOC) effect is another crucial factor when designing new magnetic materials due to its ability to interplay with spin-polarization (magnetism) of the material.[38] For example, strong SOC effect can enhance magnetocrystalline anisotropy which is important for high coercivities and energy products in ferromagnets, such as FePt[39,40]. Heavy elements with $4d/5d$ or $5p/6p$ block are normally considered to be the source of strong SOC effects, for instance, platinum and bismuth.

With these design rules in mind, we successfully designed and synthesized a new rare-earth-free ferromagnetic intermetallic compound $MnPt_5As$. Magnetic properties measurements and theoretical calculations both show consistently that $MnPt_5As$ favors ferromagnetic ordering with a saturated moment ~ 3.5 $\mu_B$ per formula. The magnetic moment in $MnPt_5As$ is theoretically proved to be strongly related to the atomic interaction between Mn and its neighboring Pt atoms. Meanwhile, $MnPt_5As$ is the first high-$T_c$ magnetic material in this structure type. The discovery of $MnPt_5As$ paves a way to more remarkable magnetic compounds with similar structures.

## Experimental Section

**Preparation of Polycrystalline MnPt$_5$As:** Polycrystalline MnPt$_5$As was synthesized by the high-temperature solid-state method. Mn powder, Pt powder, and As powder were evenly mixed and pelletized in an argon-filled glovebox due to the toxicity of arsenic.[41] The obtained pellet was placed in an alumina crucible and sealed into an evacuated silica tube (<10$^{-5}$ torr). The sealed sample was heated up to 1050 ºC at a rate of 30 ºC per hour and stayed for 2 days followed by a cooling procedure to room temperature in 5 days. The polycrystalline product is air and moisture stable.

**Phase Identification:** The phase purity was determined by powder X-ray diffraction (PXRD) conducted on a Rigaku MiniFlex 600 powder X-ray diffractometer with a Cu K$_\alpha$ radiation ($\lambda$ = 1.5406 Å, Ge monochromator). The Bragg angle measured was from 5º to 90º at a rate of 0.1 º/min with a step of 0.005º. Rietveld fitting on Fullprof was employed to obtain the weight percentage of the major phase MnPt$_5$As.[42]

**Structure Determination:** The crystal structure of MnPt$_5$As was determined by single crystal X-ray diffraction (SC-XRD). Multiple pieces of small crystals (~20×40×40 $\mu$m$^3$) were picked and measured by a Bruker Apex II diffractometer equipped with Mo radiation ($\lambda_{K\alpha}$= 0.71073 Å) at room temperature. The small crystals were mounted on a Kapton loop with glycerol. Four different crystal orientations were chosen to take the measurement with an exposure time of 10 seconds per frame. The scanning 2θ width was set to 0.5°. Direct methods and full-matrix least-squares on F$^2$ models within the *SHELXTL* package were applied to solve the structure.[43] Data acquisition was obtained *via* Bruker *SMART* software with the corrections on Lorentz and polarization effect done by *SAINT* program. Numerical absorption corrections were accomplished with *XPREP*.[44,45]

**Scanning Electron Microscope (SEM):** Crystal picture was taken using a high vacuum scanning electron microscope (SEM) (JSM-6610 LV). Samples were placed on carbon tape prior to loading into the SEM chamber and were examined at 20 kV.

**Physical Properties Measurements:** The Quantum Design Dynacool Physical Property Measurement System (PPMS) is used to measure the magnetic property and resistivity with the temperature range from 1.8 to 300 K with and without applied fields. The magnetic susceptibility is defined as $\chi$ = M/H. Here, M is the magnetization in units of emu, and H is the applied magnetic field. A standard relaxation calorimetry method was used to measure heat capacity and the data were collected in zero magnetic field between 220 K and 320 K using H-type grease. All the measurements were performed on polycrystalline samples manually selected from MnPt$_5$As.

**Electronic Structure Calculations:** The electronic and ferromagnetic structures were calculated using Tight-Binding, Linear Muffin-Tin Orbital-Atomic Spheres Approximation (TB-LMTO-ASA) with local (spin) density approximation (L(S)DA).[46–48] The empty spheres are required during the calculation with the overlap of Wigner–Seitz (WS) spheres limited to smaller than 16%. A mesh of 9×9×6 $k$-points in the first Brillouin zone (BZ) was used to perform the detailed calculations and obtain the density of states (DOS) and Crystal Orbital Hamiltonian Population (COHP) curves. The band structure, Fermi surfaces, and density of states (DOS) of MnPt$_5$As were also calculated using the WIEN2k program, which has the full-potential linearized augmented plane wave method (FP-LAPW) with local orbitals implemented.[49,50] The electron exchange-correlation potential was used to treat the electron correlation within the generalized gradient approximation, which is parameterized by Perdew et. al.[51] The conjugate gradient algorithm was applied, and the energy cutoff was set at 500 eV. Reciprocal space integrations were completed over an 8×8×4 Monkhorst-Pack $k$-points mesh.[52] With these settings, the calculated total energy converged to less than 0.1 meV per atom. The spin-orbit coupling (SOC) effects were only applied for Pt atoms. The structural lattice parameters obtained from experiments are used for both calculations.

## Results and Discussion

**Crystal Structure and Phase Determination of New Phase MnPt₅As:** To obtain the structural feature of the new phase, the single crystal of MnPt₅As was investigated. The refined results, crystallographic data and anisotropic displacement parameters are shown in Tables 1, 2 and 3. MnPt₅As crystallize in the tetragonal structure with the space group $P4/mmm$ (No. 123). Mn located at $1c$ ($C_4$ symmetry), Pt atoms are on $1a$ ($C_4$ symmetry) and $4i$ ($C_2$ symmetry) sites, and As atom occupies $1b$ ($C_4$ symmetry) sites. The partial occupancy refinements were tested, and no vacancy or mixture was discovered. A typical layered feature can be observed where the face-shared Mn@Pt₁₂ polyhedra layers are intercalated by As layers in FIG. 1$a$. The distances between the two nearest Mn atoms are 3.931 (1) Å within the $ab$-plane and 7.092 (1) Å along the $c$-axis. The lattice parameters of MnPt₅As are 0.5% and 2.9% larger in $a$ and $c$, respectively, than those in MgPt₅As. The increased lattice parameters in MnPt₅As result in increasing the bond length of Pt1-Pt1 along the c-axis from 2.756 Å in MgPt₅As to 3.129 Å in MnPt₅As. However, the Mn-Pt distance is ~2.79 Å while the Mg-Pt distance is ~2.84 Å. The shorter atomic distance of Mn-Pt indicates $Mg^{2+}$ in MgPt₅As can be replaced by the smaller $Mn^{3+}$. Moreover, we have identified a close relationship between the MnPt₅As type structure and unconventional superconductor, PuCoGa₅. Pu-centered Ga cuboctahedra (Pu@Ga₁₂) and Co-centered Ga cubes (Co@Ga₈) in PuCoGa₅ can be analogies to Mn@Pt₁₂ and As@Pt₈ in MnPt₅As, respectively.

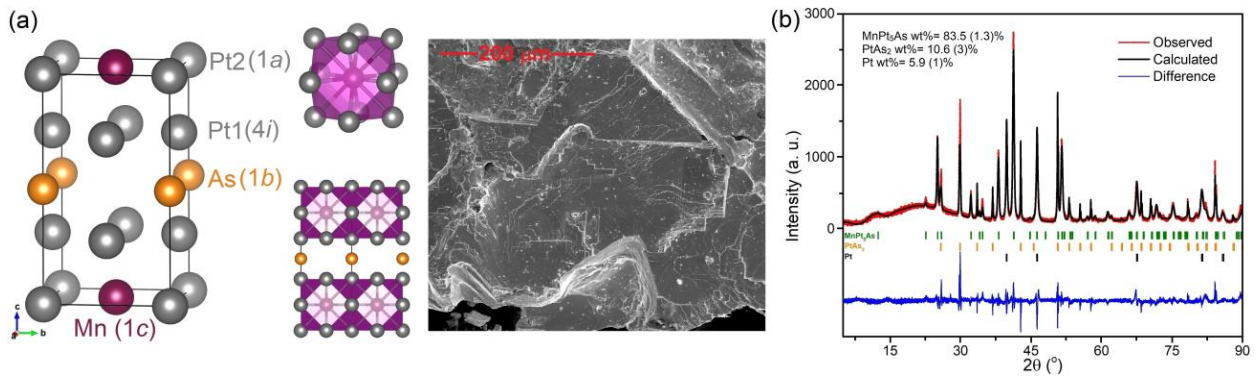

**FIG. 1** (*a*) Crystal structure of MnPt₅As determined by single crystal XRD where purple, grey and orange balls stand for Mn, Pt and As atoms, respectively. Crystal picture showing a layered feature is on the right. (*b*) Rietveld fitting of PXRD pattern from 5° to 90° including contributions from MnPt₅As, PtAs₂ and Pt phases.

By fitting the PXRD pattern shown in FIG. 1*b* with crystal structure obtained from SC-XRD, it can be realized that the majority of the product is MnPt$_5$As (83.5(1) wt%) while there are two non-magnetic impurity phases identified as PtAs$_2$ (10.6(3) wt%) and Pt (5.9(1) wt%). The fitting parameters of MnPt$_5$As phase are $\chi^2$= 3.62, $R_p$= 8.99, $R_{wp}$= 12.0 and $R_{exp}$= 6.32 and the difference between the observed and calculated patterns is acceptable while no obvious extra phases could be indexed. We also tried to refine the fourth phase, MnAs[53], as possible magnetic impurity but it showed an unreliable result due to the high R factor. Re-grinding the obtained polycrystalline samples and reannealing are not helpful to make the product purer. Instead, multiple annealing steps lower the degree of crystallization. The observed intensity of the PXRD pattern would be extremely low and most of the minor peaks cannot be identified.

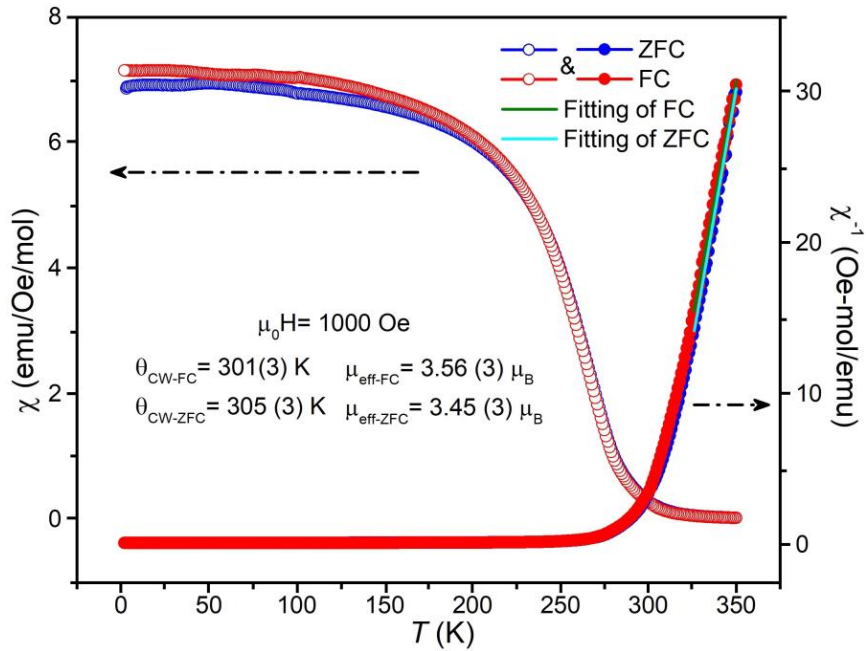

**FIG. 2.** Temperature-dependence of magnetic susceptibility (Empty Circle) and inverse magnetic susceptibility (Solid Circle) of MnPt$_5$As with ZFC (Red) and FC (Blue) methods. Green and cyan lines indicate Curie-Weiss fitting lines for FC and ZFC curves, respectively.

**Magnetic Susceptibility of Polycrystalline MnPt$_5$As:** To investigate the magnetic properties of MnPt$_5$As, polycrystalline samples of MnPt$_5$As were measured with an external magnetic field of 1000 Oe. The results of both field-cooled (FC) and zero-field cooled (ZFC) M vs T are shown in FIG. 2. As cooling down the sample, the dramatic increase of magnetic susceptibility is clearly observed near Tc ~ 300 K which implies ferromagnetic ordering. An analysis of the inverse susceptibility using Curie-Weiss (CW) law was modeled using $\chi = C/(T-\theta_{CW})$, where $\chi$ is magnetic susceptibility, C is Curie constant, and $\theta_{CW}$ is Curie-Weiss temperature. The effective moments $\mu_{eff}$ fitted at the high temperature range (325 K- 345K) ($\mu_{eff} = \sqrt{8C}$ $\mu_B$) are 3.56 (3) $\mu_B$ for FC method and 3.45 (3) $\mu_B$ for ZFC method, respectively. The fitted Curie-Weiss temperatures for both methods are 301 (3) K (FC) and 305 (3) K (ZFC), which is consistent with the ferromagnetic transition. The measurements have been conducted on different specimens to confirm the reproducibility.

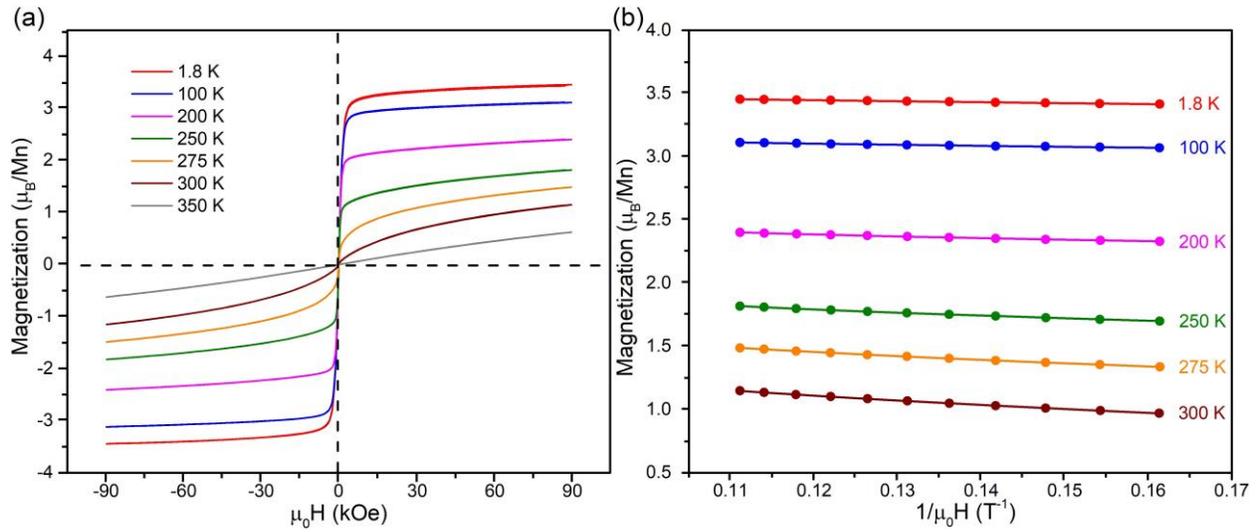

**FIG. 3. (*a*).** Hysteresis loop from -9 T to 9 T under various temperatures from 1.8 K to 350 K. (*b*). magnetization per Mn atom as a function of inverse magnetic field (1/$\mu_0$H) where the intercept of each line with linear fitting indicates the saturated magnetic moment under that temperature.

**Field-Dependent Magnetization in MnPt$_5$As:** The hysteresis loop of magnetization $M(H)$ measured with applied fields up to 90 kOe at various temperatures from 1.8 K to 350 K is shown in FIG. 3. The small coercivity (~15 Oe at 1.8 K) confirms the soft ferromagnetic character of MnPt$_5$As. FIG. 3$a$ shows the field dependences of the magnetization per Mn atom. MnPt$_5$As is magnetized to saturate at M$_s$~3.541 (1) $\mu_B$/Mn by an external magnetic field of ~0.7 T at 1.8 K. As increasing temperature, the saturated moments decrease till at 300 K and varnishes 350 K. In particular, the field dependences of $M(H)$ at 350 K investigated carefully upon both sweeping up and down of the applied field shows the paramagnetic properties. The exact saturated moments at different temperatures are determined from the intercepts in FIG. 3$b$ by linear fitting each line. The results are consistent with the magnetic susceptibility measurements. The saturated moment observed for $MnPt_5As$ is smaller than what was reported for one of the most well-known Mn-based ferromagnets, MnBi, where a saturated moment of ~4.2 $\mu_B$/Mn can be found[54]. The difference of $M_s$ between MnBi and $MnPt_5As$ might be coming from the larger Mn-Mn distance and weaker Mn-Mn magnetic interactions in $MnPt_5As$ ($d_{Mn-Mn}$ ~ 3.91Å in $MnPt_5As$; $d_{Mn-Mn}$ ~ 3.05Å in MnBi).

**Resistivity of $MnPt_5As$:** To study the electronic properties of $MnPt_5As$, the resistivity of a small single crystal of $MnPt_5As$ was picked up and measured using the 4-probe method. FIG. 4 illustrates the temperature-dependence of resistivity without an applied magnetic field. The resistivity measurement shows the metallic properties of $MnPt_5As$. An anomaly can be observed around 300 K which reflects the ferromagnetic ordering transition, similar with other ferromagnetic intermetallics, for example, $Ce_2CoGe_3$, $Ce_5Co_4Ge_{13}$ and $Y(Fe_{1-x}Co_x)_2$[55,56]. The reason for the kink observed could be that the major contribution of resistivity in $MnPt_5As$ changes from electron-phonon scattering (at a higher temperature) to electron-electron scattering (below magnetic ordering temperature) while it is possible that additional magnetoresistance contribution develops below $T_c$. A high-quality single crystal may be necessary for further investigation. The resistivity curve was fitted by using power law $\rho(T) = \rho_0 + AT^n$ where $\rho_0$ is residual resistivity due to defect scattering, A is a constant and n is an integer determined by the interaction pattern. At high-temperature (HT) region before the magnetic ordering temperature (310 K-350 K), as shown by the green line, we found that it fits well when n equals 1 which means the resistivity has mainly resulted from the collision between electron and phonon. As the temperature goes down, the phonon contribution decreases significantly while n is fitted to be 3 at the low-temperature (LT) region (4-25 K). This confirms that *s-d* electron scattering is a possible predominant mechanism at low temperature.[57] The residual resistivity $\rho_0$ for HT and LT regions was fitted to be 18.8 (1) $\mu\Omega$-cm and 3.546 (3) $\mu\Omega$-cm, respectively. The constant A was fitted to be 0.0356 (3) $\mu\Omega$-cm/K in HT range and a much smaller 3.84 (6) $\times$ 10$^{-5}$ $\mu\Omega$-cm/K$^3$.

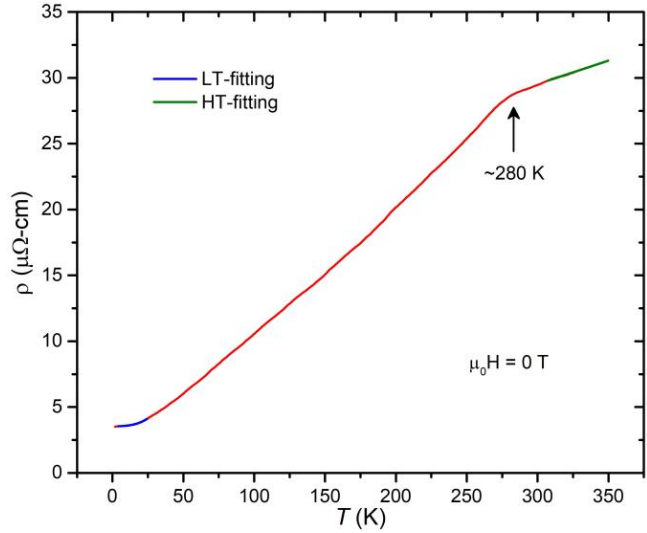

**FIG. 4.** Temperature-dependence of resistivity for MnPt$_5$As under 0 T. Green and blue lines stand for the power law fitting under high & low temperature regions.

**Heat Capacity Measurement:** The specific heat of MnPt$_5$As was measured between 220 K and 320 K, as shown in FIG. 5. A clear anomaly can be observed starting from ~300 K with the peak appearing at ~270 K, which is consistent with the result for the magnetic ordering transition temperature in magnetic susceptibility measurement. The λ–shape anomaly shown in confirms the magnetic phase transition at room temperature is intrinsic in MnPt$_5$As. Due to the lack of the phonon mode and the noisy high-temperature data, we are not able to calculate the magnetic entropy change for MnPt$_5$As.

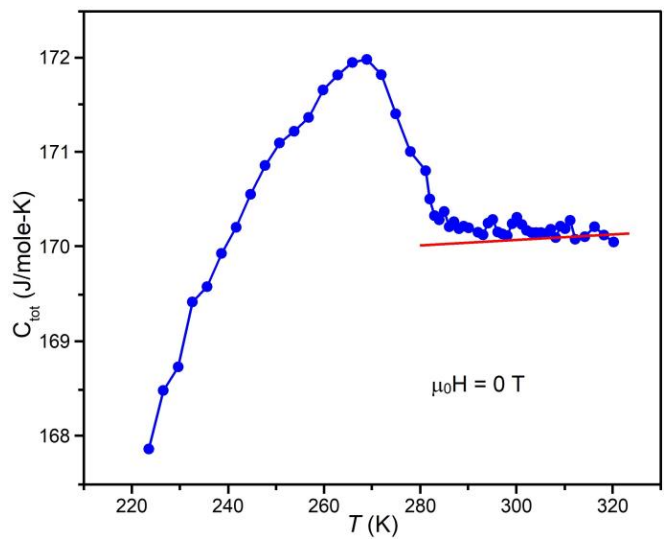

**FIG. 5.** Heat capacity measurement conducted between 220 K to 320 K with no applied magnetic field. The red line is the proposed baseline.

**Chemical Bonding and Theoretical Magnetic Structure:** To examine possible atomic interaction influences on the structural stability and magnetic properties in MnPt₅As, DOS and -COHP by TB-LMTO-ASA calculations were calculated. With the local density approximation (LDA), the corresponding DOS curve for MnPt₅As is illustrated in FIG. 6*a*, which emphasizes contributions from the Mn and Pt valence orbitals. Most of the DOS curve below -0.5eV belongs to the Pt 5*d* band; Fermi level (–1 ~ +1 eV) is dominated by a combination of Mn and Pt *d* orbital contributions, which forms a noticeably sharp and intense peak around 1eV wide. According to the corresponding -COHP curves, the wavefunctions contributing to this peak have a strong Pt-Mn and Pt-Pt antibonding character. These features of the LDA-DOS arise from structural influences on the orbital interactions in MnPt₅As. Evaluation of the Stoner condition using the Mn partial DOS gives $N(\text{E}_\text{F})I(\text{Mn}) = 6.3$; where $N(\text{E}_\text{F}) = 15.5$ eV$^{-1}$, $I(\text{Mn})$ stands for the exchange-correlation integral, 0.408 eV for Mn.[58] Thus, according to the LDA-DOS curves, MnPt₅As is susceptible toward either a possible structural distortion or toward ferromagnetism by breaking the spin degeneracy to eliminate the antibonding interactions at the Fermi level.

Applying spin polarization via the local spin density approximation (LSDA), the DOS and -COHP curves for the spin-up and spin-down wavefunctions are illustrated in FIG. 6*c-e*. It is clearly shown that the corresponding Fermi levels shift away from the peaks in the DOS curves, and closely approach the pseudogap in both majority and minority spin DOS curves. Moreover, Pt-Mn and Pt-Pt antibonding states vanish. The integration of the spin-up and spin-down contribution yields a total magnetic moment of 4.55 μ$_\text{B}$ per formula unit for MnPt₅As. Interestingly, analysis of local moments reveals the magnetic moments primarily from Mn atoms (3.95 μ$_\text{B}$) and slightly from Pt atoms (0.12 μ$_\text{B}$). The Stoner product in spin-polarized situation is 2.6 indicating that the system is favored in ferromagnetic ordering. Further investigation will be needed to determine the role of Pt atoms in Mn-Mn exchange interaction. Moreover, the majority and minority spin Mn 3*d* and Pt 5*d* bands are not fully filled at the Fermi level, which also indicates soft ferromagnetic behavior. The theoretical calculation confirms what is observed experimentally.

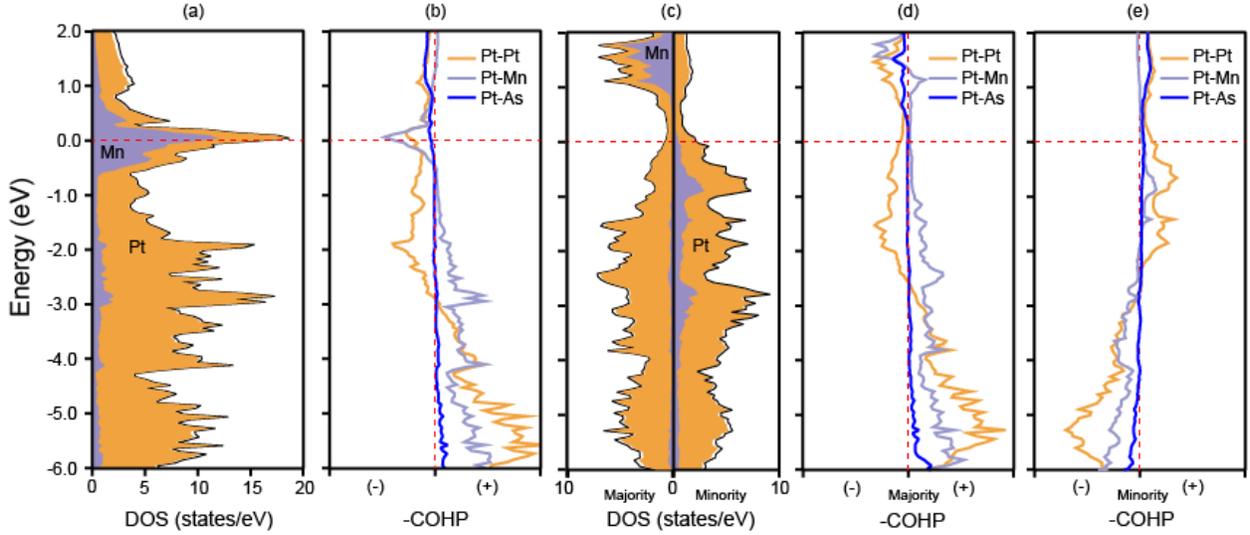

**FIG. 6.** (*a*) Partial DOS curves and (*b*) –COHP curves of MnPt$_5$As obtained from non-spin-polarization (LDA). (*c*) Partial DOS curves and (*d, e*) –COHP curves of MnPt$_5$As obtained from spin-polarization (LSDA).

**Electronic Structure Calculation:** To estimate the spin-orbit coupling (SOC) influence on the magnetic properties in MnPt$_5$As, band structure and density of states (DOS) were also calculated using the WIEN2k program. The electronic structures for MnPt$_5$As in FIG. 7*a* and *b* show great consistency of resistivity measurement that the compound is metallic. The LDA band structures with/without SOC reveal that the sharp peak in DOS involves the hybridization of Mn 3*d* and Pt 5*d* bands that are relatively flat (nearly dispersionless) near the Brillouin zone boundaries. A saddle point can be observed at Γ point near the Fermi level (E$_F$) when SOC is not applied which is broken by SOC when it is included. The SOC effect also opens the bandgap at some high-symmetry points, for instance, Γ point. The *p* orbital of P atom and *d* orbitals of Mn and Pt atoms are dominant in the energy range we showed, as can be seen in DOS figures. A van Hove singularity from *d* orbital of Mn atom can be observed near E$_F$ when SOC is not considered. However, the SOC effect splits the intensive peak in DOS near E$_F$ into two small peaks (indicated by the red arrows). Therefore, the *d* orbital from the Mn atom contributes ~50 % less at E$_F$. In magnetic materials, the tendency of being ferromagnetic ordering can be determined by using the Stoner criterion which requires a value larger than one for the term $N$(E$_F$) $I$(Mn). The term is calculated to be 8.34 without SOC and 4.41 with considering the SOC effect. The calculated Stoner parameter without SOC from the WIEN2k program is closed to the one obtained LMTO calculation. Thus, MnPt$_5$As is theoretically ferromagnetic and consistent with the experimental magnetic measurement.

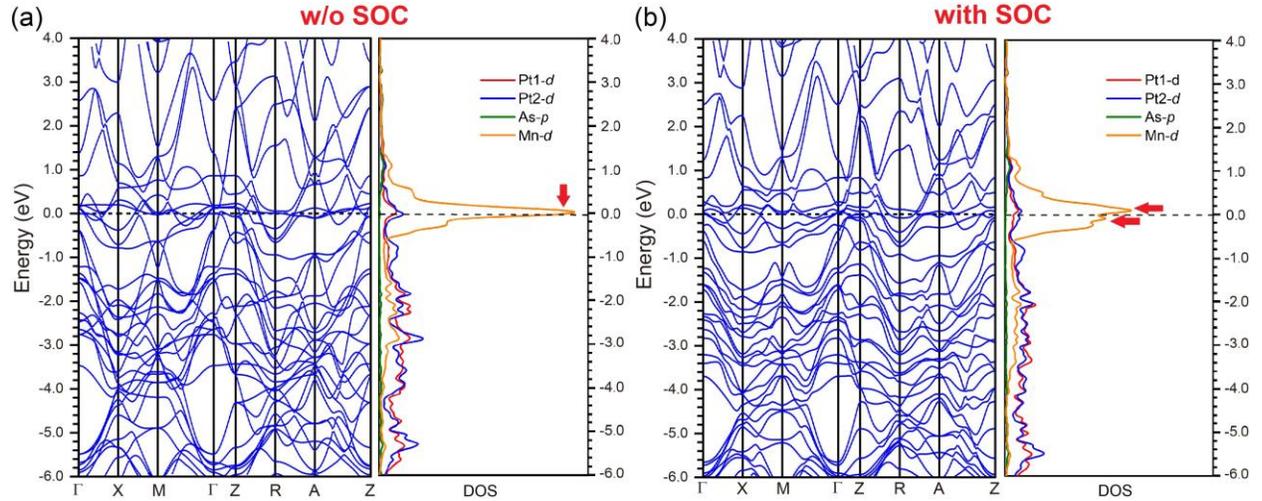

**FIG. 7.** Band structure and density of states (DOS) of MnPt$_5$As (**_a_**) without and (**_b_**) with consideration of spin-orbit coupling (SOC) effect. The red arrows indicate the splitting of van Hove singularity on DOS near E$_F$ caused by SOC effect.

Three-dimensional Fermi surface demonstrates the distribution of energy bands crossing the Fermi level in the first Brillouin Zone (BZ). To estimate the SOC effects on the electronic structure of MnPt$_5$As, the Fermi surfaces with/without SOC were generated and illustrated in FIG. 8. Intuitively, the strong SOC effect on Pt atoms indeed decreases the electronic states within the 1$^{st}$ BZ which also reflects on the density of states. SOC effect mainly influences the Fermi surface at/near the high-symmetry points which are marked in FIG. 8 except for Γ point that is at the geometric center of the first BZ. One can easily tell that by introducing SOC effect, the saddle point at R point disappears. The electronic states near the Γ point in the center of the BZ also vanish. Moreover, electronic states near A and Z points dramatically reduce after SOC effect is included.

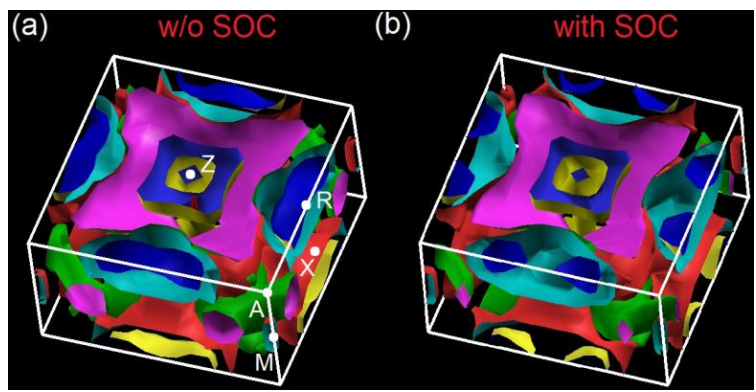

**FIG. 8.** Fermi surface of MnPt$_5$As in the first Brillouin Zone of MnPt$_5$As **(a)** without and **(b)** with spin-orbit coupling effect considered. Z, R, A, M and X indicate the position of high-symmetry points. $\Gamma$ point is at the geometric center of the first BZ.

## Conclusion

In this paper, we successfully synthesized the polycrystalline of a new rare-earth-free Pt-rich ferromagnet, MnPt$_5$As. The new material holds a layered tetragonal crystal structure with a space group of *P*4/*mmm*. Magnetic properties measurements proved that MnPt$_5$As is ferromagnetic ordered at room temperature. Resistivity and heat capacity measurements confirm the magnetic properties measurements. The theoretical electronic structure demonstrates that the Pt-Mn and Pt-Pt antibonding interactions are critical to the ferromagnetic properties in MnPt$_5$As. Moreover, the spin-orbit coupling effect of Pt atoms is essential in decreasing the density of states of Mn atom at E$_F$ by splitting the van Hove singularity at the Fermi level, thus lowering the saturated moment. Moreover, MnPt$_5$As is the first high-T$_c$ magnetic material holding this structure type. It provides a great platform to design and investigate new magnetic materials with targeted functional properties.

## Supporting Information

No supporting information is available.

## Author Information

Corresponding Author: weiweix@lsu.edu

Notes: The authors declare no competing financial interest.

## Acknowledgement

The work at LSU is supported by Beckman Young Investigator award.

**Table 1.** Single crystal structure refinement for MnPt$_5$As at 296 (2) K.

| Refined Formula | MnPt$_5$As |
|---|---|
| F.W. (g/mol) | 1105.31 |
| Space group; Z | $P\,4/mmm$; 1 |
| $a$(Å) | 3.931 (1) |
| $c$(Å) | 7.092 (1) |
| V (Å$^3$) | 109.58 (5) |
| θ range (º) | 2.872-36.301 |
| No. reflections; $R_{int}$ | 1489; 0.0408 |
| No. independent reflections | 200 |
| No. parameters | 11 |
| $R_1$: $\omega R_2$ ($I>2\delta(I)$) | 0.0275; 0.0730 |
| Goodness of fit | 1.165 |
| Diffraction peak and hole (e$^-$/ Å$^3$) | 4.430; -5.266 |

**Table 2.** Atomic coordinates and equivalent isotropic displacement parameters for MnPt$_5$As at 296 (2) K. ($U_{eq}$ is defined as one-third of the trace of the orthogonalized U$_{ij}$ tensor (Å$^2$))

| Atom | Wyckoff. | Occ. | $x$ | $y$ | $z$ | $U_{eq}$ |
|---|---|---|---|---|---|---|
| Pt1 | $4i$ | 1 | 0 | ½ | 0.2794 (1) | 0.0063 (2) |
| Pt2 | $1a$ | 1 | 0 | 0 | 0 | 0.0058 (2) |
| As3 | $1b$ | 1 | 0 | 0 | ½ | 0.0066 (5) |
| Mn4 | $1c$ | 1 | ½ | ½ | 0 | 0.0073 (8) |

**Table 3.** Anisotropic thermal displacements from MnPt$_5$As.

| Atom | U11 | U22 | U33 | U23 | U13 | U12 |
|---|---|---|---|---|---|---|
| Pt1 | 0.0064 (2) | 0.0050 (2) | 0.0075 (3) | 0 | 0 | 0 |
| Pt2 | 0.0060 (3) | 0.0060 (3) | 0.0052 (4) | 0 | 0 | 0 |
| As3 | 0.0069 (7) | 0.0069 (7) | 0.006 (1) | 0 | 0 | 0 |
| Mn4 | 0.008 (1) | 0.008 (1) | 0.006 (2) | 0 | 0 | 0 |

**For table of content only**

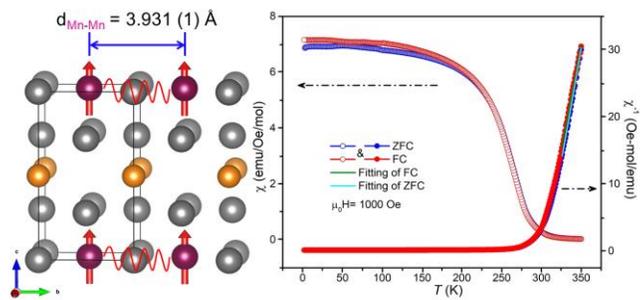